\documentclass{edp-conf}
\usepackage{graphicx}
\usepackage{a4wide}
\usepackage{amssymb} 

\newcommand{\apj} {ApJ}
\newcommand{\mnras} {MNRAS}
\newcommand{\aap} {A\&A}

\newcommand{\fig}[3]{
      \begin{figure}[tbp]
	\resizebox{\hsize}{!}{\includegraphics  {#1}}
	\caption{#2}
	\label{#3}
        \end{figure} }
\newcommand{\eqn} [1] {
\begin{equation} 
#1 
\end{equation}}

\newcommand{\inv} {\frac {1}}

\begin{document}

\TitreGlobal{SF2A 2002}

\title{Consequences of the  non gaussian character of the stochastic excitation for solar-type oscillations.} 

\author{Samadi R.$^{1,4}$}
\author{Nordlund A.$^{2}$} 
\author{Stein R.F.$^{3}$}
\author{Goupil M.-J.$^{4}$} 
\author{Roxburgh I.$^{1,}$}

\address{Astronomy Unit, Queen Mary, University of London, London, UK.}
\address{Niels Bohr Institute for Astronomy, Physics and Geophysics, Copenhagen, Denmark.}
\address{Michigan State University, USA}
\address{LESIA, Observatoire de Paris, Meudon, France.}

\runningtitle{Consequences of the non gaussian character of the stochastic excitation}
\setcounter{page}{237}
\index{Author1, A.}
\index{Author2, B.}
\index{Author3, C.}

\maketitle
\begin{abstract} 
Our recent study, based on a 3D numerical simulation, suggested that the stochastic excitation of solar $p$-modes has a non gaussian nature
and a non gaussian model was  proposed which results in a good agreement between the computed rate $P$ at which $p$-modes are excited and the solar seismic observational constraints.

In the present work, we study some consequences of this non-gaussian model on the rate $P$  at which $p$-modes are excited in  intermediate mass stars ($1 M_\odot \lesssim M \lesssim  2 M_\odot $). 
The non gaussian model   changes substantially the spectrum of $P$ at high frequency  compared with the commonly assumed  gaussian model.
The largest effects on $P(\nu)$ are found for stars in the mass range $\sim 1.5-1.6~M_\odot$.
They are found large enough that observatins in solar-like oscillating stars observed with space based seismology missions (e.g. COROT) will be able to confirm or reject the gaussian nature of stochastic excitation.
\end{abstract}

%
\section{Introduction}

Excitation of solar-type oscillations results from turbulent movements in the upper convective zone of intermediate mass stars ($1 M_\odot \lesssim M < 2 M_\odot $).
The rate $P(\nu)$ at which a given mode with frequency $\nu$ is excited crucially depends on the time averaged and \emph{dynamic} properties of the turbulent medium (Samadi 2001).

Different models for  stochastic excitation have been proposed by several authors (e.g. Goldreich \& Keeley 1977; Balmforth 1992; Goldreich et al. 1994; Samadi \& Goupil 2001).
The model of Samadi \& Goupil (2001, Paper~I hereafter,  see also Samadi (2001)) takes into account the frequency dependence $\chi_k(\nu)$ of the correlation product of the turbulent velocity field.
This last quantity characterizes the dynamic properties of the turbulence.
This excitation model offers then the advantage to test different 
hypothesis concerning the dynamic properties of the turbulent medium.

A gaussian function is usally assumed for $\chi_k(\nu)$ (e.g. Goldreich \& Keeley 1977).
However recently, using 3D numerical simulation of the upper part of the solar convective zone, Samadi et al. (2002)  have shown that the gaussian model indeed does not correctly model  $\chi_k(\nu)$ in the frequency range where the acoustic energy injected into the solar $p$-modes is important  ($\nu \simeq 2 - 4$~mHz). One must consider an additional non-gaussian component for $\chi_k(\nu)$ to reproduce its behavior. Computed values  of $P$ obtained with this non-gaussian component reproduce better the solar seismic observations.

In the present work we study the consequences of this non-gaussian component for the  stochastic excitation of $p$-modes in  intermediate mass stars. 
Sect.~1 describes the stochastic excitation model we consider and sumarizes Samadi et al. (2002)'s results which establish the non gaussian character of the stochastic excitation.
In Sect.~2 we compute $P$ for different stellar models with masses in the mass range $1 M_\odot \lesssim M \lesssim  2 M_\odot $. In the computation we assume two hypotheses for $\chi_k(\nu)$ : the  gaussian function and a non gaussian function as inferred in Samadi et al. (2002) from a 3D simulation.

\section{The non gaussian character of the stochastic excitation}

The theoretical expression in Paper~I for the excitation rate $P$ can be written in a schematic form as :
\eqn{
P(\nu) \propto \int d{\rm m} \, \int d{\rm r} \, d\tau \, \, \vec \xi_{\rm r} \, .  \,< \vec S \, \vec S >({\rm r},\tau) \, .  \,\vec \xi_{\rm r}
\label{eqn:P}
}
where $\rm r$ and $\tau$ are related to the local turbulence : $\rm r$ is the distance between two given points in the turbulent medium and $\tau$ is the time correlation  between the two distant points ($\tau$ is associated with the eddy lifetime which is typically much shorter than the oscillations lifetime). 
In Eq.~(\ref{eqn:P}),  $\xi_{\rm r}$ denotes the radial mode eigenfunction and $<\vec S \, \vec S>$ the correlation product of the excitation source versus ${\rm r}$ and $\tau$. The excitation source ($\vec S$) originates from the turbulent Reynolds stress and the advection of the turbulent entropy fluctuations by the turbulent motions.
$<\vec S \, \vec S>$  is expressed in terms of the turbulent kinetic energy spectrum $E(k,\nu)$ and the spectrum of the entropy fluctuations  $E_s(k, \nu)$ where $k$ is the wavenumber of a given turbulent element.
Further $E(k,\nu)$ is  split  into a spatial component $E(k)$ and a frequency component $\chi_k(\nu)$ according to Stein (1967) as
$ E( k,\nu) =E(  k) \, \chi_k(\nu) $
The same decomposition is assumed for $E_s(k, \nu)$.

In the calculation of  $P(\nu)$, a gaussian shape is  usually assumed for $\chi_k(\nu)$ (e.g. Goldreich \& Keeley 1977).
This assumption is equivalent to suppose that two distant points in the turbulent medium are uncorrelated.
Such  Gaussian function  (GF hereafter) has the form
\eqn{
\chi_k^{\rm GF} (\nu ) = \inv  { \nu_k \, \sqrt{\pi}}  e^{-(\nu / \nu_k)^2} \; ,
\label{eqn:GF}
}  
where $\nu_k$ is the line-width at half maximum.

With the help of a 3D numerical simulation of the upper part of the solar convective zone, it is found in Samadi et al. (2002) that  the GF does not reproduced correctly the frequency dependence of $\chi_k$  inferred from the 3D simulation.
In contrat, $\chi_k$ is better modeled by the Gaussian  plus an Exponential function  (GEF hereafter) as: 
\eqn{
\chi_k^{\rm GEF}  (\nu ) = {1 \over 2 } \, \left (  \inv  { \nu_k \, \sqrt{\pi}}  e^{-(\nu / \nu_k)^2} + \frac{1}{2 \nu_k}  e ^{-| \nu/\nu_k | } \right ) \;  .
\label{eqn:GEF}
}


\section{Consequences for solar-like oscillating stars}

We compute  several  stellar models with masses $M=1., 1.2,1.5,1.65,1.85, 2.~M_\odot$ and with the same age ($650~My$). Calculations are carried out with the CESAM evolutionary code in which the convective heat flux is computed according to the classical 
mixing-length theory by B\"ohm-Vitense (1958). The eigenfunctions are obtained from the 
adiabatic FILOU pulsation code by Tran Minh \& Leon (1995). The input physics are similar to those of Samadi et al. (2001) except for  the atmosphere which  is computed with the Eddington classical gray atmosphere for sake of simplicity.

\fig{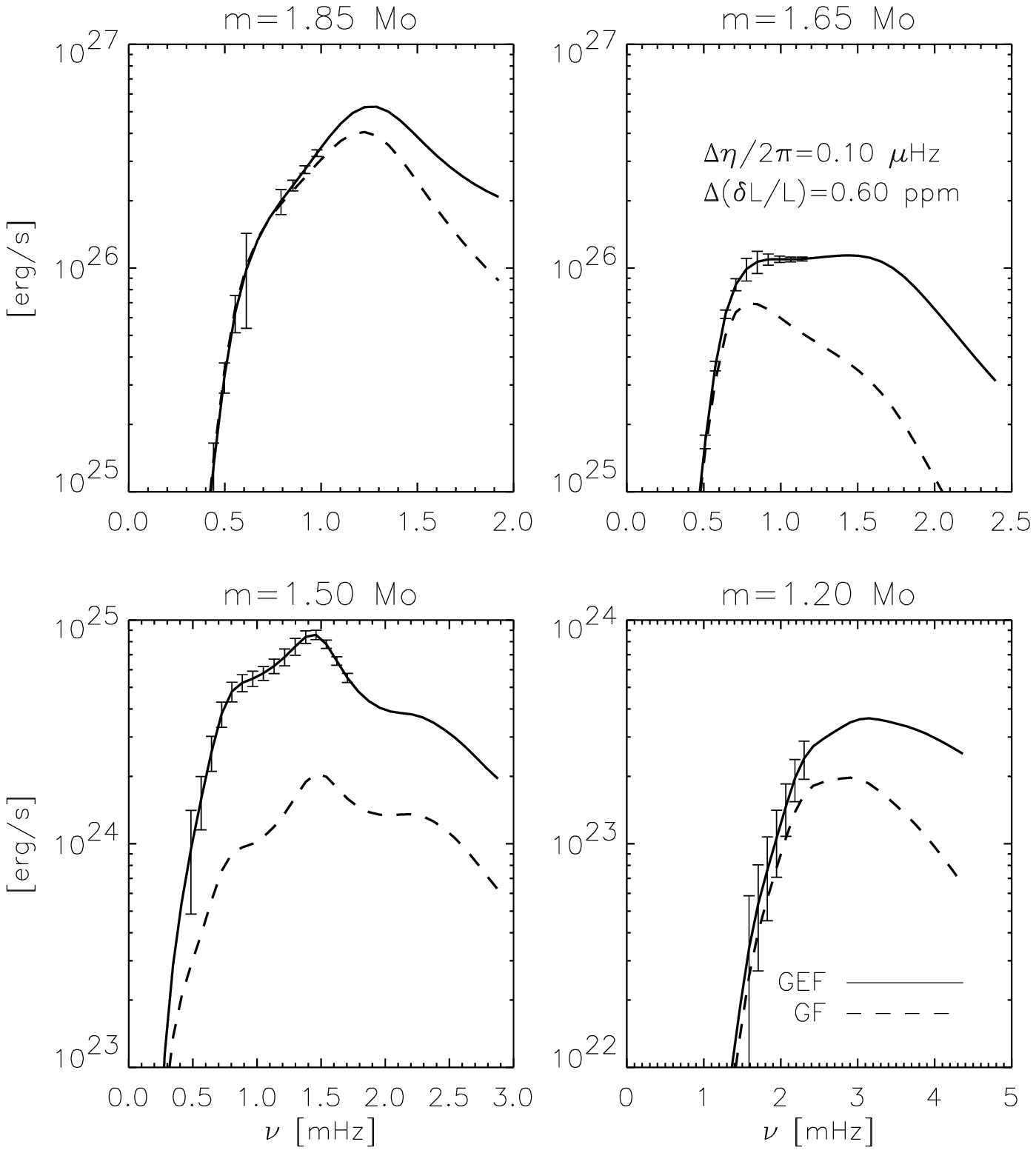}{The excitation rate $P$ at which stellar $p$-modes are excited is computed  for several  stellar models with masses $M=1.2,1.5,1.65,1.85~M_\odot$. Two different analytical functions are assumed for $\chi_k(\nu)$~: the GF (dashed curve, Eq.~\ref{eqn:GF}) and the GEF (solid curve, Eq.~\ref{eqn:GEF}). Vertical error bars $\Delta P$ are estimated in the manner of Samadi et al. (2001).  The assumed accuracy in the measurement of the mode damping rates ($\eta$)  and the mode  relative luminosity fluctuations ($\delta L/L$) are based   on the expected performances of the COROT mission (Baglin et al. 1998) (i.e. $\Delta \eta /2\pi = 0.1~\mu$Hz and $\Delta \delta L/L= 0.6$~ppm). Values of $\eta$ are calculated in the manner of Houdek et al. (1999). 
}{fig:cmp_pow_gexp_m100_m200}

We compute $P$ for each the stellar model according to Eq.~(\ref{eqn:P}).
In the calculation, we assume successively for $\chi_k(\nu)$~:  the gaussian model (GF, Eq.~\ref{eqn:GF}) and the non gaussian model (GEF, Eq.~\ref{eqn:GEF}). 
A scaling factor is introduced such that for the solar model the maximum in $P$ matches the solar seismic constraints by Chaplin et al. (1998).
Results are presented  in Fig.~\ref{fig:cmp_pow_gexp_m100_m200}.

For all stellar models, the largest differences in $P$ between calculations performed with the GF and those performed with the GEF are obtained for $\nu \gtrsim \nu_{\rm max}$ where $\nu_{\rm max}$ is the frequency at which $P$ is maximum. This feature is a direct a consequence of the difference in the frequency dependence of  the GF and the GEF.
Indeed, the GEF decreases with $\nu$ slower than the GF does for $\nu \gtrsim \nu_k$. 
Moreover  $\nu_{\rm max}$ scales approximatively as $\bar \nu_k$ where $\bar \nu_k$ is the characteristic frequency associated with the largest energy-bearing eddy located at the top of superadiabatic (this is the region where the excitation is the largest).
The eddies who excite  the most the modes have their characteristic frequency $\nu_k \simeq \bar \nu_k$. Consequently for $\nu \gtrsim  \nu_{\rm max} $ a larger amount of acoustic energy is injected with the GEF than the GF.

The largest effects in $P$  of the dynamic model adopted for $\chi_k$ are obtained for the $M=1.5~M_\odot$ and $M=1.65~M_\odot$ stellar models~: Let  $\Delta P$ be the difference in $P$ between calculations performed with the GF and those performed with the GEF.
$\Delta P$ is found larger for these two models  than for the other stellar models. 
Moreover  $\Delta P$  takes large values over a larger range in  $ \nu / \nu_c$ where   $\nu_c$ is the cut-off frequency ($\nu_c$  delimits the extend of frequency domain where modes have a significant amplitude).
These features are due to the fact that the characteristic turbulent Mach number, $M_t$, is the highest  for stars with mass $\sim 1.5-1.6~M_\odot$.
Indeed, $\Delta P_{\rm max}$ , the maximum in  $\Delta P$, scales approximatively as $\bar \nu_k^3 \, \Delta \chi_k $ where $ \Delta \chi_k \equiv  \| \chi_k^{\rm GEF}   - \chi_k^{\rm GF}   \|_{\nu=\bar \nu_k} $.
As   $ \Delta \chi_k $ scales as ${\bar \nu_k}^{-1}$ and $\bar \nu_k$ scales approximatively as $M_t \,  \nu_c$, $\Delta P_{\rm max}$  increases then as $M_t^2$.
Moreover, let $\Delta \nu$ be the frequency domain where  $ \| \Delta P - \Delta P_{\rm max} \| < \epsilon \, \Delta P_{\rm max} $ where $\epsilon$ is arbitrarily fixed to a value $\ll 1$. 
In other words $\Delta \nu$  defines  a  domain where  $\Delta P$ is ``large''.
One shows that $d^2 \Delta \chi_k / d \nu^2$ scales as ${\bar \nu_k}^{-3}$ and then  $d^2 \Delta P / d \nu^2 \propto {\bar \nu_k}^{-2} \, \Delta P_{\rm max}$.
As a consequence  $  \| \Delta P - \Delta P_{\rm max}    \| \approx{ 1 \over 2} \,   {\bar \nu_k}^{-2}  \, \Delta \nu ^2 \, \Delta P_{\rm max}$. 
The domain  $\Delta \nu/\nu_c$  then  scales  as $\sqrt{\epsilon} \, \bar \nu_k / \nu_c \propto M_t$.

\section{Conclusion}

The present work illustrates that a non gaussian model of stochastic excitation 
substantially changes the shape and the magnitude of the calculated excitation spectrum $P$ of the oscillations in intermediate mass star more massive than the Sun.
Differences in  $P$ between computation carried out with the non gaussian excitation model and this  carried out with the common gaussian model are found larger for the frequency modes above the frequency $\nu_{\rm max}$ at which $P$ peaks.
The largest effects are expected for  stars in the mass range  $\sim 1.5-1.6~M_\odot$.
They are found large enough to be discriminate from space based experiences with performances equivalent or higher than those of the COROT intrument (Baglin et al. 1998).


\bibliographystyle{aa}



\end{document}